# On Possibility of Using High-$T_c$ Ceramic-Superconductor as Junction-less Transistor towards Nano-miniaturization

R.C. Gupta*, Ruchi Gupta** and Sanjay Gupta***

## Abstract

High-Tc Type-II ceramic-superconductor at temperature $T < T_c$, under presence of magnetic-field B becomes non-superconducting if B exceeds a critical value $Bc_2$. Thus at $T < T_c$, by application/absence of critical magnetic- field as a controlling device, these non-superconducting/superconductor states can be achieved for current-flow to two corresponding states of block/pass or off/on or 0/1. Thus it appears that there is a possibility of a new breed of transistors purely with high-Tc Type-II ceramic-superconductor; compact and without junctions & complexities. The proposed ceramic-superconductor-transistor (CST) seems in-principle to work well for switching purpose, but its use could also be extended for other electronic/computer devices too. The CST, being junction-less thus diffusion-less, could possibly be packed more closely (at nano-level) than the semi-conductor devices which has a limitation due to diffusion-layer-overlapping. A similar superconductor-device named Cryotron was invented at MIT half-a-century ago, but could not survive against semiconductor. CST is a rebirth of cryotron in different disguise & in new perspective.

## Introduction

The driving force behind every technological-revolution is *'material'*. Behind industrial-revolution it is *steel* and behind electronic-revolution it is *semiconductor*. What is the material for next technological-revolution ? Is it *superconductor* ? There are many useful & important applications of the superconductors and here is one more possible application of it. Most of electronic/computer circuits/devices based on transistors are made of semiconductors, the SQUIDs, however, make use of Joshepson-Junction (JJ) of two superconductors with an insulating layer in-between; but it is proposed (as an extension of cryotron) here that bare ceramic-superconductor could also be used as a new breed of junction-less transistor and may even prove to be better in absence of junction defects/problems.

In digital circuits for electronic/computer devices, the transistor usually is used as a two-state device or switch. The state of a transistor can be used to set a voltage on the conducting path/channel to be either high or low, representing a binary one or zero. Logical and arithmetic functions are implemented in a circuit built using transistors as switches. A transistor's other function is amplification of small input signal to high output signal through a controller-parameter, such as through current in BJT and through voltage in FET(MOS). All BJT, FET, MOSFET including the most popular CMOS for VLSI are basically semi-conductor junction devices [1-3] whereas a few others such as SQUIDs use superconductor junctions [4,5].

The authors do not claim originality of the idea of using superconductor as transistor, credit goes to Prof. Dudley Buck of RLE, MIT (USA) for his then-famous invention [6] of 'Cryotron' using low-Tc metallic-superconductors; the authors however suggest use of high-Tc ceramic-superconductors in new perspective of nano-technology. Prof. Buck was most recognized for the development of 'cryotron', a super-conductive magnetically controlled gating device that was hailed as a revolutionary component for miniaturizing the room-sized computers of 1950s. His technical paper "The Cryotron – A Superconductive Computer Compound" was recognized and awarded with the Browder Thomson Medal of the Institute of Radio Engineers (now IEEE) in 1957. The cryotron is a switch that operates using superconductors, and works on the principle that the magnetic-field destroys superconductivity. The cryotron [7] is a piece of tantalum wrapped with a coil of niobium placed in liquid helium bath. When the current flows through the tantalum-wire it is super-conducting, but when a current flows through the niobium-coil a magnetic-field is produced which destroys the superconductivity of tantalum thus stopping the current-flow in tantalum. Historically, first came the vacuum-tube, then the semiconductor-transistor; the cryotron was destined to spark another revolution in electronics, but could not, as the semi-conductors (at room-temperature) succeeded over metallic-superconductors (at very low temperature). The demise of cryotron was probably due to rapid developments in semiconductor devices and due to untimely death of Dudley Buck in 1959. The authors believe that a rebirth of 'cryotron' can take place in new shape & size as 'CST'; with the advent of modern developments in field of high-Tc ceramic-superconductors, nano-technology & manufacturing-techniques.
.
A switch/valve (or transistor) usually has 3 terminals for electron-flow: (i) in (Source), (ii) out (Drain) and (iii) control barrier/base (Gate). The signal can be changed through control terminal signal from 'this-state' to 'that-state'. The 'this/that' combination may be considered as pass/no-pass, channel open/closed, on/off, high/low, 1/0 etc; or in quantum computing (qubit) this/that combination is spin-up/spin-down [3].

---

*Dr.R.C.Gupta, Prof. & Head, Mechanical Engg. Dept., Institute of Engineering & Technology (IET),
  Lucknow, India  (E-mail: rcg_iet(at)hotmail.com).
** Ruchi Gupta, Software Engineer, Cisco System, CA, USA  (ruchig(at)stanfordalumni.org).
*** Sanjay Gupta, Staff Engineer, Vitris Technologies, Ca, USA (sanjaygupta01(at)yahoo.com).

In the present-paper it is to question ( & answer) the possibility: 'if a ceramic-superconductor *without junction* can be used as transistor specially as switching device and if so then *how* and *what is* 'this/that' combination for it and what are the 3 terminals and what is the signal with which to control the current-flow' ? Also discussed are possible difficulties & prospects and implementation of the proposed CST and its future.

**Superconductors**

It is known that certain materials [4-5] show superconductivity (zero resistance) below certain critical - temperature Tc ; application of certain amount of magnetic-field Bc , however, destroys the superconductivity (Meissner-effect). Based on the penetration & response of magnetic-field, there are two types of superconductors: Type-I and Type-II [4,5]. With the application of increasing magnetic-field, Type-I loses its superconductivity abruptly whereas Type-II loses its superconductivity gradually. Type-I superconductors have very low critical temperature Tc (below which the material is super-conducting) of the order of $10^o$ K whereas Type-II superconductors are having comparatively high Tc upto of the order of $150^o$ K. Type-I superconductors are usually pure metal-conductors. Type-II superconductors are either (a) conducting metal alloys & compounds or even (b) non-conducting ceramics. Type-II-(a) metal alloys & compounds have Tc of the order of $30^o$ K or even higher whereas Type-II-(b) ceramics, usually called High-Tc superconductors, now have Tc of the order of $150^o$ K much higher than boiling point ($77^o$ K) of liquid nitrogen - which is easily available & affordable so convenient to use.

Superconductor have several uses such as to make high-strength-magnets for various applications ranging from MLT (bullet-train) to NMR (MRI). The other applications include use of Josephson-Junction (JJ) for *s*uper-conducting-*qu*antum-*i*nterface-*d*evice*s* (SQUIDs) for use as a sensitive magnetometer and high speed fluctuating (switching) device. In JJ(SQUIDs) two Type-I superconductors have a junction with a thin insulating layer in-between through which the Cooper-electron-pair can tunnel, it works on overlapping of electron's wave-functions φ on both sides [4,5]. Cryotron is another feasible application discussed earlier; and CST as explained in this paper could be the next brighter application of superconductor as futuristic transistor.

It may be noted that Type-I metals and Type-II-(a) metal alloys/compounds superconducting materials are always at-least good-conductors whether they are in superconducting-state or not; whereas Type-II-(b) ceramic superconducting material is excellent-conductor in superconducting-state but not-so-good-conductor (comparatively, insulator) in non-superconducting-state. There was a big-boost to the development of super-conductor research with the surprise discovery of ceramic (insulator) being newer & better superconductor by German scientists Bednorz and Muller [8] in 1986. Then-after world-wide efforts are on, to get even higher Tc ceramic-superconductor with a desire to achieve Tc up to room-temperature.

**Possibility of Using High-Tc Ceramic-Superconductor as Junction-less Transistor**

It is well known that in presence of applied magnetic field B at T < Tc, if the field exceeds a certain critical value $Bc_2$ which depends on the superconductor-material & its temperature, its superconductivity disappears altogether[4,5] . If B is switched off to zero or reduced to B < $Bc_1$ , it reverts back to superconducting-state. If $Bc_1$ < B < $Bc_2$ it is in mixed-state i.e., partially superconducting and partially non-superconducting [4,5] i.e., in-between superconductor & non-superconductor meaning partial-conductor.

Thus if applied magnetic field B > $Bc_2$ , the High-Tc Type-II-(b) Ceramic-Superconductor becomes non-super-conducting. This switch-over phenomenon, from superconductor to non-superconducting with application critical magnetic-field, can be used as transistor-switch for current-flow as pass/no-pass due to material flipping as superconductor/non-superconducting. All superconductors including High-Tc Type-II-(b) 'Ceramic-Superconductors' can work as good 'transistor-switch' due to its dual material characteristics 'of superconductor as well as of non-superconducting', depending on its conduction state of high/low as pass/no-pass or 1/0 . The ceramic-superconductor, appears also to work as *smart-material*, is capable to *sense* the amount of applied magnetic-field B through variations in its conductivity with the field ($Bc_1$ < B < $Bc_2$).

So if a small bare wire or strip or rectangular-dot of this High-Tc Ceramic-Superconductor is taken (at T < Tc) with one-end as 'Source' & the other-end as 'Drain' for electron-flow and that if magnetic-field B is applied across it such that B > $Bc_2$ then the superconductor will become non-superconducting. As mentioned / asked about the 'terminals' and 'this/that' earlier in the last-paragraph of Introduction-section, it to say that if High-Tc Ceramic-Superconductor is used as junction-less Transistor then the 3 terminals are : (i) one-end as S (Source), (ii) other-end as D (Drain) & (iii) middle-region as G (Gate) for magnetic-field B and 'this/that' combination is current-flow 'pass/no-pass' due to 'superconductor becoming non-superconducting' with the application of critical magnetic-field on Gate-region to control the current-flow.

Normally semi-conductor devices (e.g., switch) require an external-load (resistance) for voltage-drop, for the switch to work as *inverter*. The resistance of the superconductor (in non-superconducting state) can itself act as if it is *the* load in superconductor-devices (Cryotron & CST). Interestingly, Logic Gates can be easily constructed with *no-external-load*, and considering control signal as B & resulting voltage signal towards end-S of the superconductor as V. Taking applied-voltage (at D) $V_D$ = 1 & critical-magnetic-field $Bc_2$ = 1 units, and considering 'presence / absence' of the resulting voltage as high/low or '1 / 0'; the result is that - if B =0, V = 1 and else if B = 1, V = 0 ; the proposed ceramic-superconductor-transistor thus works as '*inverter*' switch. Logic Gates such as NAND or NOR can be built using two such *load-less inverters* in series or in parallel respectively. Further digital circuits can be logically built on combination of various Logic Gates.

The 'Ceramic-Superconductor Transistor' is briefly named here as CST. The CST seems to work well as far as 'switching' is concerned, and can well be used to build digital circuits. But using it for input-signal 'amplification', it apparently doesn't seem to work; because the control signal, is not directly the small current at Base as in BJT or small voltage at Gate as in FET, which does amplification therein . Control-signal in CST is the small change (possibly due to small change of current or voltage at Gate) in magnetic field B which does not directly *enter* into the circuit; so it is not clear that how this control-signal, which is external & outside to the circuit, can control or magnify/vary the input-signal of current or voltage in the circuit. One suggestion to achieve the control of circuit-current through external magnetic-field B, is to vary B with the help of control-signal at Gate to increase/decrease the total magnetic-field B in a range such that $Bc_1$ < B < $Bc_2$ ; i.e., keeping the ceramic-superconductor in mixed-state of partially superconducting & partially non-superconducting; *varying the conductivity with control-signal through changes in B within the mixed-region*. CST– electronics yet to be fully-developed would be significantly different and may open new avenues /applications.

The symbol (as shown in Fig.1) of the new breed of transistor - Ceramic Superconductor Transistor (CST) is suggested as follows with a special plus sign. The vertical thick-line (S-D) shows the ceramic-superconductor; it is made thick to differentiate from the thin circuit-wires (not shown in the figure) to be connected to S , D & G. The horizontal dotted-line with an arrow shows the control-signal i.e., variable vector magnetic-field B; the line is dotted and only apparently-intersecting with the thick-line because in fact the field B is produced outside the circuit but surrounds all-around over the superconductor and can change its property through control-signal at Gate. It may be noted that the CST *itself* is junction-less.

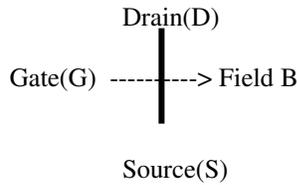

**Fig.1 Suggested Symbol for the Proposed Ceramic-Superconductor-Transistor (CST)**

      For comparison and clarity purpose, a tabular presentation is given as follows, which indicates that use of the ceramic-superconductor-transistor (CST) is almost as viable & feasible as other devices. It may be noted that materials for all BJT, FET(MOS) & Qubit devices are semi-conductors [1-3] mostly diffusion-based with-junctions, whereas for the last three categories the materials are superconductors [4,5]. But it may be noted that JJ is with-junctions whereas cryotron and CST are junction-less. In fact, in a way, the proposed CST is a rebirth of forgotten-cryotron but now with high-Tc ceramic-superconductor and with new design & manufacturing.

**Table. 1 Transistor-types and its Details**

| Transistor-type (& material) | Terminals for electron-flow | | | Control with change in | 'this/that' combination |
|---|---|---|---|---|---|
| | In(Source) | Out(Drain) | Control(Gate) | | |
| BJT (say, n-p-n) (semiconductor) | n | n | p | current | thin/thick barrier -of depletion layer |
| FET (MOS) (semiconductor) | source(S) | drain(D) | gate(G) | voltage | open/closed -channel |
| Qubit (semiconductor) | one-end | other-end | middle-region | voltage | spin-up/spin-down -of electron |
| SQUID (JJ) (low-Tc metallic -superconductor) | super- conductor-1 | super- conductor-2 | thin insulating-layer | voltage | change of current (phase) |
| Cryotron (low-Tc metallic -superconductor) | one-end | other-end | concentric outer-coil | magnetic-field (through change in I) | superconductive / non-superconductive for current flow |
| CST (high Tc ceramic superconductor) | one-end | other-end | middle-region printed-ring on reverse-side of chip-board | magnetic-field (through change in I) | superconductor / non-superconducting for current-flow |

## Possible Difficulties, Prospects, Implementation and Future of Ceramic-superconductor as Junction-less Printable Transistor

**Difficulties:**

The difficulties for it could be (i) working at cryogenic temperature, (ii) limitations for micro or nano miniaturization, (iii) need for confinement of magnetic-field etc. Most of the existing nano-electronic-devices [9] have also to work at cryogenic temperature and with easy availability of liquid nitrogen, it should not be much problem for this too. Micro & nano miniaturization is gradual process, can be achieved in due course with modern fabrication-technologies. Confinement & implementation of magnetic-field for micro-devices could be a real big problem, but possibly with some shielding/focusing / scanning techniques with special field-gun or with some novel method (as suggested under the heading: Implementation) it could be made possible eventually.

**Prospects:**

As mentioned above the difficulties do exist but could be overcome in due course; on the other side the prospects for this new-breed of ceramic-superconductor are no less. This proposed ceramic-superconductor device as transistor, does not have any junction at all and neither have too much terminal connections & layers but, is just the bare superconductor wire/strip itself with external magnetic-field on it. So, whole device(s) can be printed on one-side of chip-board as printed-circuit with high-Tc type-II ceramic super-conducting material, a step forward in nano-technology based VLSI / GLSI. Thus the chip made with ceramic-superconductor can be more densely packed, without any impairment of devices due to overlapping of depletion-layer & leakage etc. the usual undesirable features of compact semi-conductor devices, because there would be *no* junction or depletion-layer & its associative-defects in *ceramic-superconductor-transistors* an interesting & promising feature of the *CST*.

**Implementation:**

The authors suggest a novel method for implementing (applying) magnetic field B on Gates(G) by printing superconducting rings (of only $350^o$, open $10^o$ for current-connections) on the reverse-side of the board and also the circuitry to pass the control-signal (current) through these rings to produce the magnetic field to apply on. On one-side of the chip-board are printed (i) the superconducting nano-strips (one-end as Source-S & other-end as Drain-D); whereas on the other-side of the board are printed (ii) the superconducting rings (as Gate-G), with current-circuitry printed on both side with inter-connections if needed. It could also be possible to print/implement both-these i.e., (i) the superconducting strips & (ii) superconducting rings on the *same side* of the chip but with oxide ($SiO_2$) layer in-between, somewhat analogous to MOSFET with metal connections as required. Alternatives to the usual-printing of all-these (IC) through the conventional deposition & etching / photolithographic techniques are: electron-beam evaporation, laser ablation techniques etc., or could be combination of various computer-controlled fabrication/ manufacturing techniques for electronic-devices / computer-chips [1-2, 11].

**Future:**

Famous physicist Prof. Feynman once predicted that 'there is plenty of room at the bottom'. Another famous physicist Prof. Michio Kaku predicts *technology-change* in his famous book 'Visions' [10] that 'beyond 2020, because of size-limit of silicon-chip technology, eventually we will be forced to invent new technologies whose potentials are largely unexplored and untested. …Eventually, the reign of the semi-conductor microprocessors will end, and new types of quantum devices will take over'. Professor Kaku further adds that 'in the computer industry, it takes roughly fifteen years, on the average, from conception of an idea to its entering into marketplace'. He also expects 'by 2020, *point one barrier* ( **.**1 μ wavelength of ultraviolet light for photo-etching) to end'. In the next fifteen years, with current trends, it is quite likely to get ceramic-superconductor at room temperature; the printable (IC) superconductor-devices / chips may then (by 2020) replace the diffusion-based semiconductor devices. Now it is up to the Scientists, Engineers, Institutions, Organizations, Laboratories and Industries to take-up the proposed-CST forward; CST could be the transistor of the future !

**Cryotron versus CST**

Although CST, in principle, is a rebirth of cryotron but there are several difference, stated as follows for clarity:

| | **Cryotron** | **CST** | **Remarks** |
|---|---|---|---|
| **- Introduced by in year** | Invented by Dudley Buck in 1957 | Proposed by Guptas in 2005 | Rebirth of cryotron as CST after half-a-century ! |
| **- Material** | Metallic-superconductor | Ceramic-superconductor | Ceramic-superconductor found in 1986 & onwards |
| **- Type** | Type-I superconductor | Type-II(b)*superconductor | *ceramic |
| **- Critical Temp.** | Very low Tc (<10$^o$K) | High Tc (>100$^o$K)* | *attained upto 150$^o$K and future prospect upto Room Temp |
| **- Cooling Agent** | Costly liquid Helium (4$^o$K) | Cheap liquid Nitrogen (77$^o$K) | |
| **- Principle** | Loses superconductivity with B abruptly if B > Bc | Loses superconductivity with B gradually if B > Bc$_2$ > Bc$_1$ | |
| **- Application** | Suitable for switch | Suitable for switch and also as amplifier etc* | *versatile transistor |
| **- Junction ( ? )** | Junction-less | Junction-less | whereas semi-conductor devices are made & based on junctions |
| **- Construction** | Tantalum wire wrapped with Niobium coil | Ceramic-superconductor tiny strips & rings printed on opposite-sides of chip-board | |
| **- Magnetic-field B produced by** | 3-D Coil | 2-D* Ring | *flat-thin |
| **- Direction of B-field-application** | Axial on Ta wire | Transverse* on printed superconductor -strips | *perpendicular |
| **- Manufacturing** | Vacuum evaporation…. | Vacuum evaporation / Printing…. | |
| **- Miniaturization** | Possibly Macro-size | Possibly Nano-size* | *CSTcould be more suitable than semi-conductor devices- in which junction diffusion thickness overlap causes limitation of size |
| **- Future prospects** | Fair | Bright* | *CST could be the transistor of the future |

**Conclusions**

It seems possible to develop a new breed of transistor using high-Tc Type-II ceramic-superconductor (kept at T < Tc) which becomes non-superconducting if a certain magnetic-field (B > $Bc_2$) is applied, thus making current-flow from 'pass' state to 'no-pass' state. This new proposed transistor-device (CST) is supposed to work well as switching device without any junctions/layers, flipping to condition (achieved/ controlled) through effective-change in external magnetic-field. It can also work for other purposes of transistor. The proposed ceramic-superconductor-transistor (CST) can find new avenues & applications, but its theory is yet to be fully-understood & developed as the CST-electronics would be quite different; for CST it would not be simply electronics but magneto-electronics ! CST would be load-less and junction-less thus diffusion-less printable- transistor controlled by magnetic-field. Practicality of a device is a different question which is usually resolved in due course of time; but it appears that the proposed CST, though speculative it may seem, at-least works in principle. The revival of the forgotten-cryotron as CST seems to be vital for nano-electronics; *CST could be the transistor of the future*.

**Acknowledgement**

The authors thank Dr. G.P. Gupta, Professor & Head and Dr. B.Das Asstt.Professor, Physics Department of Lucknow University for valuable discussions & advice. The authors also like to give thanks to: Dr. D.S. Chauhan, Dr. A.K. Khare, Dr. N. Malaviya, Dr. S.R.P Sinha, Dr. Sanjay Mishra, Prof. H.N. Gupta, Prof. K.K. Srivastva, Prof. Arun Mittal, Er. Atul Kumar, Er. R.K.Gautam, Er. A.K.Gupta for useful comments, and to: Sanjiv, Chhavi, Veena & Shefali for cooperation. Thanks are also due to the author's institutions / companies, UPTU, AICTE, World-Bank (TEQIP), Central & State Governments for their direct / indirect grants & facilities.